# Research on The Cultivation Path of Craftsman Spirit in Higher Vocational Education Based on Survey Data

*Yufei Xie, Jing Cui, Mengdie Wang*
*Jiangxi College of Applied Technology, Ganzhou 341000, Jiangxi, China*

Abstract: With the development of China's economy and society, the importance of "craftsman's spirit" has become more and more prominent. As the main educational institution for training technical talents, higher vocational colleges vigorously promote the exploration of the cultivation path of craftsman spirit in higher vocational education, which provides new ideas and directions for the reform and development of higher vocational education, and is the fundamental need of the national innovation driven development strategy. Based on the questionnaire survey of vocational students in a certain range, this paper analyzes the problems existing in the cultivation path of craftsman spirit in Higher Vocational Education from multiple levels and the countermeasures.
Key words: Higher vocational education; Craftsman spirit; Training path

## 1.INTRODUCTION

In 2016, at the fourth session of the 12th National People's Congress, the premier pointed out in the government work report that "enterprises should be encouraged to carry out personalized customization and flexible production, cultivate the spirit of craftsmanship to improve quality, improve quality and create brands." "Craftsman spirit" appears in the government work report, which is refreshing. Some media have listed it as "ten new words" to interpret it. In order to meet the needs of the rapid development of modern economic society and make China into a manufacturing power, we should cultivate technical and skilled talents to meet the needs of modern economic and social development, and vigorously promote the exploration of the cultivation path of craftsmanship spirit in higher vocational education, which provides new ideas and directions for the reform and development of higher vocational education, and is the fundamental need of the national innovation driven development strategy.

## 2. THE ESSENCE AND CONNOTATION OF CRAFTSMAN SPIRIT

Craftsman spirit is not only a kind of professional spirit, but also a specific embodiment of professional ethics, professional ability and professional quality. It is a kind of professional value orientation and behavior performance of practitioners. It can be specifically understood as the key concepts of dedication, lean, focus and innovation. Dedication is a kind of professional spirit state of devotion, conscientiousness and responsibility produced by practitioners based on their awe and love for their profession; lean is to strive for perfection, which is the professional quality of practitioners concentrating on each and every process, striving for perfection and pursuing the ultimate; concentration is the spirit of patience, persistence and persistence with firm mind and focusing on details Innovation is the innovation connotation of pursuing breakthrough and innovation. [2] These qualities and spirits are the aspects and excellent qualities that craftsmen must possess, and they are also the core of craftsman spirit. Craftsmanship spirit is an important yardstick of social civilization and progress, the spiritual source of making in China, the brand capital of enterprise competitive development, and the moral guidance of employees' personal growth. The spirit of craftsman is the spirit of creation, the spirit of quality and the spirit of customer first.

## 3. INVESTIGATE AND ANALYZE THE CURRENT SITUATION OF CRAFTSMAN SPIRIT TRAINING IN HIGHER VOCATIONAL EDUCATION

Through a random questionnaire survey of 600 students in a higher vocational college, the general understanding of the current situation of craftsman spirit cultivation in higher vocational education is obtained. The survey results and countermeasures are analyzed as follows:

3.1 What is the current school's emphasis on the cultivation of craftsmanship spirit?

| Survey topics | Theoretical level | Practical level |
|---|---|---|
| Which aspect does the current school craftsman spirit cultivation emphasize? | 70.8% | 29.2% |

Through the feedback of question 1 and reviewing the previous studies, we can find that they have provided many valuable suggestions and programs for cultivating craftsmanship spirit in higher vocational education, but there are still some deficiencies. For example, the current research focus of scholars focuses on the theoretical point of view, and many vocational colleges are also partial to the theoretical level of the cultivation of craftsmanship spirit. For example, 70.8% of the students think that the training focus of the school is the theoretical level, only 29.2% of the students think that the training focus of the school is the practical level, forming a sharp contrast. Therefore, there are many training methods in higher vocational colleges Some scholars believe that cultural discrimination, social change, education system defects, improper school running mode and poor combination of work and learning are the reasons for the lack of craftsman spirit cultivation in Higher Vocational Education in China.

We need to transform the theory into something of practical significance through practice, and require us to suit the medicine to the case, put forward specific plans, solve the existing problems, and provide suitable soil for the cultivation of craftsman spirit in higher vocational education, lay a good foundation and create a good atmosphere, So as to promote the integration of craftsman spirit and higher vocational education, improve the teaching quality and level of higher vocational education, and cultivate high-quality talents and technical talents.

3.2How well do vocational college students understand craftsman spirit?

| Survey topics | understand | know a little | Don't know much | Do not understand |
|---|---|---|---|---|
| Do you know the spirit of craftsmanship? | 27.8% | 49.8% | 17.5% | 4.8% |

The results of question 2 are as follows: 27.8% of the students "understand" the craftsman spirit, 49.8% of the students "know a little bit" about the craftsman spirit, while only 27.8% of the students have a good understanding of the craftsman spirit. From the survey results, we can see that the students in higher vocational colleges have a "one-sided" understanding of craftsman spirit, and their understanding of its connotation needs to be improved. This requires that our higher vocational colleges should strengthen the connotation of students' craftsmanship spirit, so that students can have a comprehensive and profound understanding of craftsman spirit.

3.3Will you take the initiative to learn and understand the spirit of craftsmanship?

| Survey topics | often | occasionally | never |
|---|---|---|---|
| Will you take the initiative to learn and understand the spirit of craftsmanship? | 19.7% | 72.2% | 8.2% |

According to the survey results of question 3, compared with high school students, students themselves have fewer courses and lighter learning tasks. As a result, some students who lack self-control are easy to develop lazy and negative behavior habits. 19.7% of the students often take the initiative to learn and understand the craftsmanship spirit, while the occasional students account for 72.2%. Moreover, some students are addicted to online games Since pull out all day holding a mobile phone to play games, swipe video, and in learning is not active, no motivation, usually do not work hard, only to pass the exam, lack of perseverance. In the face of setbacks and adversity, giving up these behaviors are the internal opposite of the pursuit of craftsmanship spirit, and are the factors that lead to students' one-sided understanding of craftsman spirit.

This requires students to correct their own attitude,

increase their learning enthusiasm and initiative, take the initiative to explore knowledge, strengthen self-control, constantly improve self-consciousness, and apply the intrinsic quality and spirit of craftsman spirit to study and life. For example, in the face of homework, internship and homework can not be dealt with hastily. Instead, we should strive for improvement and never give up until we fail to achieve the goal. We should also have the spirit of specialized research. This can not only improve our professional skills, but also better understand the craftsman spirit and eliminate "one sidedness". The school should integrate the craftsmanship spirit into the campus culture, through the related speech contest, exhibition and other campus cultural activities. To create a good atmosphere and environment for students to learn craftsmanship spirit not only enriches the students' amateur cultural life, but also broadens their knowledge, enhances their practical ability, and plays a positive role in promoting the formation of their craftsman spirit. For example, holding essay competition or speech activities related to craftsman spirit can improve students' quality and consciousness. The pursuit of quality and consciousness is also the important connotation of craftsman spirit.

Campus culture can make students further contact and learn craftsmanship spirit with rich forms. Teachers need teachers to put the craftsman spirit throughout the teaching process, simplify the complex theory, increase teaching interaction to solve problems for students, so that students can understand the comprehensive and profound craftsman spirit in the classroom.

3.4 In the cultivation of craftsmanship spirit, do you think the teaching staff is important?

| Survey topics | Very important | Generally, it is optional | unimportance |
|---|---|---|---|
| In the cultivation of craftsmanship spirit, do you think the teaching team is important | 90.7% | 7.3% | 2.0% |

In higher vocational education, it is self-evident that the importance of cultivating the craftsman spirit of teachers is self-evident. 90.7% of the students think that the teacher team is very important in the cultivation of craftsman spirit, and 2.0% of the students think that the teacher team is very important in the cultivation of craftsmanship spirit. Because teachers are the guide for students to learn craftsman spirit, many students' understanding of craftsman spirit mostly comes from teachers, but at present, many higher vocational colleges do not have special training institutions and the training mechanism is not sound enough. Therefore, the establishment of a long-term and effective teacher training mechanism is the prerequisite and inevitable requirement for the development of craftsman spirit in higher vocational education. To let the craftsman spirit take root in higher vocational colleges, teachers should first master the craftsman spirit and become a craftsman type teacher. They need to have professional dedication, improve teaching, be meticulous, diligent and selfless dedication Quality, take the initiative to shoulder the responsibility of inheriting and developing craftsman spirit, combine teaching methods with the connotation of craftsman spirit, and combine details and quality as craftsmen require for products, so as to improve teaching quality and quality, broaden and emancipate one's thoughts and horizons. It is bound to be less impetuous, more prudent, less opportunistic, more down-to-earth and less More focused, more utilitarian. Such a team of teachers can better cultivate high-quality talents and technical talents.

Secondly, higher vocational colleges should also make full use of their own resources to improve the training of teachers. For example, regular lectures on learning craftsmanship spirit should be held, and some "old craftsmen" who are excellent in education should be invited to share their teaching experience and experience. Secondly, they should organize to participate in some production links that can reflect the craftsman spirit, such as micro carving, fine carving and rice grain Calligraphy and painting, etc., is to carve characters and draw on small objects, and use knives instead of pens when carving. It is an ancient technology, and it organizes personal practice to experience the process of improving. We can also organize to watch some film and television materials, strengthen teachers' understanding of craftsmanship spirit, make the teaching team full of atmosphere and environment for learning craftsman spirit, so as to influence students imperceptibly, improve the innovative spirit and concept of teachers, enthusiasm and practical ability of teaching methods, and use the old to bring new concept, from more experienced, teaching experience Rich teachers lead the teachers who have just begun to work to improve the overall quality of the teaching team, and constantly improve the training mechanism of the teaching team.

3.5 Do you think it is necessary to innovate the curriculum for the cultivation of craftsmanship spirit?

| Survey topics | be necessary | There is no need |
|---|---|---|
| Do you think it is necessary to innovate the curriculum of craftsmanship cultivation | 92% | 2% |

Due to the cultivation of craftsmanship spirit, the curriculum still follows the traditional theoretical education. 92% of the students think that the innovation of the curriculum is very necessary. The lack of innovative curriculum will make it difficult for students to feel the fun of learning in the classroom teaching, reduce students' learning enthusiasm, which is not conducive to the development of craftsman spirit in higher vocational education, which requires teachers in higher vocational colleges, To keep pace with the times, keep up with the trend, improve their own sense of innovation, according to the characteristics of their own disciplines, targeted development of effective education courses, individualized teaching, traditional curriculum innovation and development to inject vitality into the curriculum, can use the network platform, new media means of combination, bold attempt, continuous innovation of curriculum and education methods, at the same time pay attention to improve teaching Interest in learning, to make students interested in interest, interest is the best teacher, it will drive students to take the initiative to learn, so that the spirit of craftsmanship in the hearts of students. For example, combining traditional teaching methods with new media means, at present, students have a mobile phone, and mobile phone is an indispensable communication medium and an important way for students to understand external information. We can make full use of wechat, microblog, post bar and other apps, use app as teaching carrier, and release some knowledge of craftsmanship spirit on microblog and wechat platform Or the article can also shoot some small videos, so that students can use mobile phones to learn the spirit of craftsmanship anytime and anywhere. It can also guide interested students to exchange and discuss on platforms such as post bar or microblog, hold award-winning Q & A and other activities, and attach importance to the mode of combining online and offline.

Teachers can also make use of live online teaching and offline learning to consolidate the way students love to explain the course, which is easier for students to accept, so as to ensure the effectiveness of the course. For example, with the help of the current advanced VR technology, the VR technology and boring theory are combined to make the boring theory vivid and develop some puzzle VR games. For example, students majoring in automobile maintenance develop a VR game of engine maintenance troubleshooting and engine assembly, so that students can personally understand the connotation and application effect of craftsman spirit in the game, so as to have a deeper understanding of craftsman spirit, stimulate students' enthusiasm for participation and learning, improve professional skills, and enhance their hands-on ability, autonomy and creativity, To enhance the majority of students on the craftsman spirit training course knowledge of a new height, truly achieve the maximum effect of teaching courses.

3.6 Do you think the school's teaching infrastructure is perfect?

| Survey topics | perfect | Imperfections need to be strengthened |
|---|---|---|
| Do you think the school's teaching infrastructure is perfect? | 47.7% | 52.3% |

The teaching infrastructure of a school determines the quality and level of teaching. It can be seen from question 6 that many higher vocational colleges do not attach great importance to the cultivation of craftsmanship spirit, and the infrastructure needs to be strengthened. As a result, the cultivation of craftsmanship spirit has no special practice operation site, which makes students have poor practical ability and can not personally experience the connotation of craftsman spirit. Therefore, higher vocational colleges should have special venues for craftsman spirit cultivation. For example, workshops can become the incubation base for students to learn craftsmanship spirit. First of all, higher vocational colleges should pay more attention to the cultivation of craftsmanship spirit, increase the investment in infrastructure for the cultivation of craftsmanship spirit, establish different workshops according to different majors and open them to students free of charge, and update workshop equipment in time. The school can also cooperate with enterprises to improve the infrastructure with the funds and equipment of enterprises, and enterprises can also obtain Directional training of talents to achieve a win-win situation, the enterprise's equipment and production process technology directly moved to the "workshop", so that students in the school learning process can zero distance contact with the enterprise's environment, equipment and production process technology, so that the students not only have a strong professional ability, but also because they are familiar with the enterprise, they can work directly without internship, which improves students' graduation After the employment rate and professional competitiveness also make students adapt to the future work environment to enhance the ability.

4.CONCLUSION

To sum up, through the survey results, we can know that in the cultivation path of craftsman spirit in higher vocational education, at the level of students, curriculum, teachers and schools, there are some key problems, such as students' one-sided understanding of craftsman spirit, lack of innovation in curriculum teaching mode, imperfect teacher team training mechanism, and imperfect teaching infrastructure. Higher vocational colleges should speed up the transformation of cultivation focus from theoretical level to practical level, so that students can have a comprehensive understanding and understanding of craftsmanship spirit, start from themselves, strengthen the publicity of campus culture, create a strong atmosphere, establish and improve the training mechanism of teachers, create a group of craftsman teachers, and use new technologies such as information network and VR technology to keep up with the trend of the times and boldly try to innovate continuously In order to realize the organic integration and sustainable development of craftsmanship spirit in higher vocational education, the curriculum drives students to study actively, improve and follow up the infrastructure, and create workshops to temper "craftsmen".